\documentclass[%
 reprint,
superscriptaddress,
longbibliography,
 amsmath,amssymb,
 aps,
floatfix
]{revtex4-1}

\usepackage[T1]{fontenc}
\usepackage{graphicx}
\usepackage{dcolumn}
\usepackage{bm}
\usepackage{amsmath}
\usepackage{lmodern}
\usepackage{mathtools}

\usepackage{xcolor}
\usepackage{graphicx}
\usepackage{dcolumn}
\usepackage{bm}
\usepackage[hidelinks]{hyperref}

\usepackage[normalem]{ulem}

\hypersetup{
    colorlinks,
    linkcolor= [rgb]{0.1804,0.1882,0.5725},
    citecolor=[rgb]{0.1804,0.1882,0.5725},
    urlcolor=[rgb]{0.1804,0.1882,0.5725}
}

\usepackage[T1]{fontenc}
\usepackage{graphicx}
\usepackage{dcolumn}
\usepackage{bm}
\usepackage{amsmath}
\usepackage{lmodern}
\usepackage{mathtools}

\usepackage{upgreek}
\usepackage{chemformula}
\usepackage{tabularx}
\usepackage{makecell}

\usepackage{xcolor}
\usepackage{siunitx}
\usepackage{dcolumn}

\usepackage{tcolorbox}
\usepackage{colortbl}
\definecolor{cor}{RGB}{249,230,209}
\definecolor{Cor}{RGB}{252,242,232}
\definecolor{COR}{RGB}{254,252,249}

\usepackage{float}

\newcommand*{\figref}[2][]{%
  \hyperref[{#2}]{%
    Fig.\,\ref*{#2}%
    \ifx\\#1\\%
    \else
      #1%
    \fi
  }%
}

\usepackage{capt-of}

\begin{document}

\title{Exciton-polaritons in multilayer WSe\textsubscript{2} in a planar microcavity}

\author{M.\,Kr\'ol}
\author{K.\,\,Rechci\'nska}
\author{K.\,Nogajewski}
\author{M.\,Grzeszczyk}
\author{K.\,\L{}empicka}
\author{R.\,Mirek}
\author{S.\,Piotrowska}
\affiliation{Institute of Experimental Physics, Faculty of Physics, University of Warsaw, ul.~Pasteura 5, PL-02-093 Warsaw, Poland}
\author{K.\,Watanabe}
\author{T.\,Taniguchi}
\affiliation{National Institute for Materials Science, Tsukuba, Ibaraki, 305-0044, Japan}
\author{M.\,R.\,Molas}
\affiliation{Institute of Experimental Physics, Faculty of Physics, University of Warsaw, ul.~Pasteura 5, PL-02-093 Warsaw, Poland}
\author{M.\,Potemski}
\affiliation{Institute of Experimental Physics, Faculty of Physics, University of Warsaw, ul.~Pasteura 5, PL-02-093 Warsaw, Poland}
\affiliation{Laboratoire National des Champs Magn\'etiques Intenses, CNRS-UGA-UPS-INSA-EMFL, Grenoble, France}
\author{J.\,Szczytko}
\author{B.\,Pi\k{e}tka}
\email{Barbara.Pietka@fuw.edu.pl}
\affiliation{Institute of Experimental Physics, Faculty of Physics, University of Warsaw, ul.~Pasteura 5, PL-02-093 Warsaw, Poland}

\begin{abstract}
Due to high binding energy and oscillator strength, excitons in thin flakes of transition metal dichalcogenides constitute a perfect foundation for realizing a strongly coupled light-matter system. In this paper we investigate mono- and few-layer WSe\textsubscript{2} flakes encapsulated in hexagonal boron nitride and incorporated into a planar dielectric cavity. We use an open cavity design which provides tunability of the cavity mode energy by as much as 150\,meV. We observe a strong coupling regime between the cavity photons and the neutral excitons in direct-bandgap monolayer WSe\textsubscript{2}, as well as in few-layer WSe\textsubscript{2} flakes exhibiting indirect bandgap. We discuss the dependence of the exciton's oscillator strength and resonance linewidth on the number of layers and predict the exciton-photon coupling strength.
\end{abstract}

\maketitle

\section{Introduction}

The possibility to engineer multilayered structures using two-dimensional (2D) van der Waals materials allows to construct new artificial crystals of designed functionalities \cite{Geim_Nature2013} with strong application potential in photonics and opto-electronics \cite{Mak_NatPhoton2016}, including, especially, the areas of photo-detectors, ultrathin-film photovoltaic devices, light emitting diodes and lasers \cite{Britnell_Science2013,Koppens_NatNanotechnol2014,Ahn_2DMater2016,Pu_AdvMater2018,Binder2019}.

Optical properties of monolayer (ML) transition metal dichalcogenides (TMDs) are widely explored as they are dominated by particularly strong exciton resonances stable up to room temperatures \cite{Wang_RevModPhys2018,Koperski_Nanophotonics2017}. This is a consequence of their direct bandgap in contrast to few-layer and bulk forms that exhibit indirect optical transitions \cite{Splendiani_NanoLett2010,Mak_PRL2010,molasNanoscale}. These properties make ML TMDs interesting for laser light sources  \cite{Wu_Nature2015,Ye_NatPhoton2015,Li_NatNanotechnol2017} and light-matter interactions which demand narrow-resonance oscillators \cite{Basov_Science2016,Schneider_2018}. For ML, the lowest energy transition occurs between the band-edge states at the $K^\pm$ points of the Brillouin zone, which are mostly composed of the transition metal atom orbitals: $d_{x^2-y^2}\pm {\rm i} d_{xy}$ orbitals in the valence band (VB) and $d_{z^2}$ orbitals in the conduction band (CB). Starting from the bilayer, an indirect transition between the band-edge states at the $\Gamma$ point in the VB and the Q point in the CB appears, as those states are partially composed of $p_z$ orbitals of the chalcogen atoms, which significantly overlap between the layers \cite{Kormanyos_2DMater2015,Liu_ChemSocRev2015,Lindlau_NatCom2018}.

For a long time the indirect bandgap of multilayer TMDs was considered  a drawback in the construction of opto-electronic devices due to low emission rate and poor quantum efficiency. However, compared to the ML, multilayers of TMDs have also several advantages as they possess larger optical density of states and exhibit higher absorbance as well as longer exciton lifetimes~\cite{Li_Nanoscale2018}. It is also possible to tune the bandgap and alter the carrier type in TMD-based FET devices through the control over the TMD film thickness \cite{Pudasaini_NanoResearch2018}.

\begin{figure*}
\centering
\includegraphics[width=15.3cm]{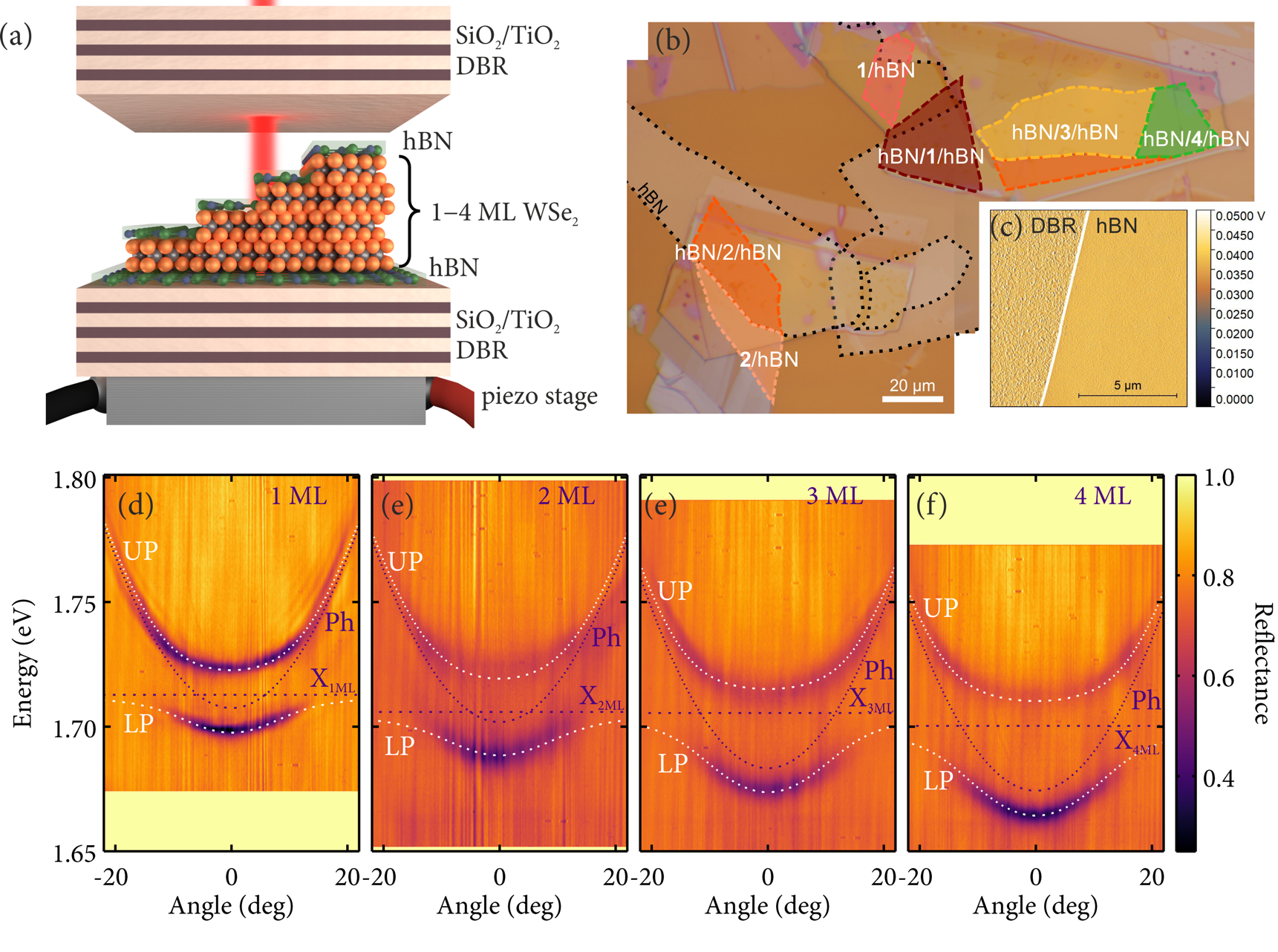}
\caption{(a) Scheme of a tunable cavity with few-layer flakes of WSe\textsubscript{2} encapsulated in hBN. (b) Microscopic image of investigated WSe\textsubscript{2}/hBN and hBN/WSe\textsubscript{2}/hBN heterostructures deposited onto a distributed Bragg reflector (DBR). (c) Amplitude image obtained with the help of an atomic force microscope operated in the tapping mode, revealing the difference in surface roughness between the DBR and hBN. (d--f) Angle-resolved reflectance spectra of hBN/WSe\textsubscript{2}/hBN heterostructures with 1\,ML- to 4\,ML-thick WSe\textsubscript{2} flakes shown in (b) embedded in a dielectric cavity. The white dashed lines mark fitted upper and lower polariton dispersion relations given by a coupled oscillators model with the energies of the uncoupled excitonic and photonic modes drawn with the purple dashed lines.}
\label{im:Fig1}
\end{figure*}

In this work, we explore the high optical absorbance of multilayer TMDs and demonstrate the strong light-matter coupling between the cavity photons and neutral excitons in multilayer WSe\textsubscript{2} placed in a tunable, planar optical resonator. Multiple works have already demonstrated the strong coupling regime for excitons in TMD monolayers incorporated into various types of planar optical cavities \cite{Lundt_2DMater2016}: dielectric \cite{Liu_NatPhoton2014}, tunable \cite{Dufferwiel_NatCommun2018,Flatten_NatComm2017}, metallic \cite{Sun_NatPhoton2017}, semiconductor--metallic \cite{Waldherr_NatCommun2018}, but all of them were entirely focused on direct-bandgap ML flakes or MLs separated by thin hexagonal boron nitride (hBN) flakes \cite{Dufferwiel_NatComm2015}. 

The strong light-matter coupling conditions in multilayer TMDs have been recently achieved in different systems with plasmonic cavities of ultra-small volume \cite{Kleemann_NatComm2017,Stuehrenberg_NanoLett2018} and in open plasmonic cavities formed by periodic arrays of metallic nanoparticles \cite{Wang2019}. In that case, however, the resonant electric field is polarized perpendicularly to the flakes' plane and nearly perpendicularly to the excitons' dipole moment which is mostly in-plane oriented \cite{Schuller_NatNanotech2013}. Here we report on the observation of the strong coupling regime between photons confined in a planar cavity, where the electric field is polarized along the main axis of the excitons' dipole moment, with neutral excitons in ML as well as few-layer-thick WSe\textsubscript{2} flakes. Starting from the bilayer WSe\textsubscript{2} the bandgap becomes indirect, but the absorption at direct excitonic resonance is efficient enough to allow for creation of strongly coupled light--matter states. 
The exciton--photon coupling strength reaches even higher values for multilayers than for  the ML due to increased thickness of the TMD material. 
Our findings are well reproduced by calculations done within the framework of a transfer matrix method, where the parameters of the excitonic resonances in WSe\textsubscript{2} are determined based on the analysis of reflectance spectra of the flakes deposited directly onto the bottom distributed Bragg reflector (DBR).

\section{Results}

\begin{figure*}
\centering
\includegraphics[width=15.3cm]{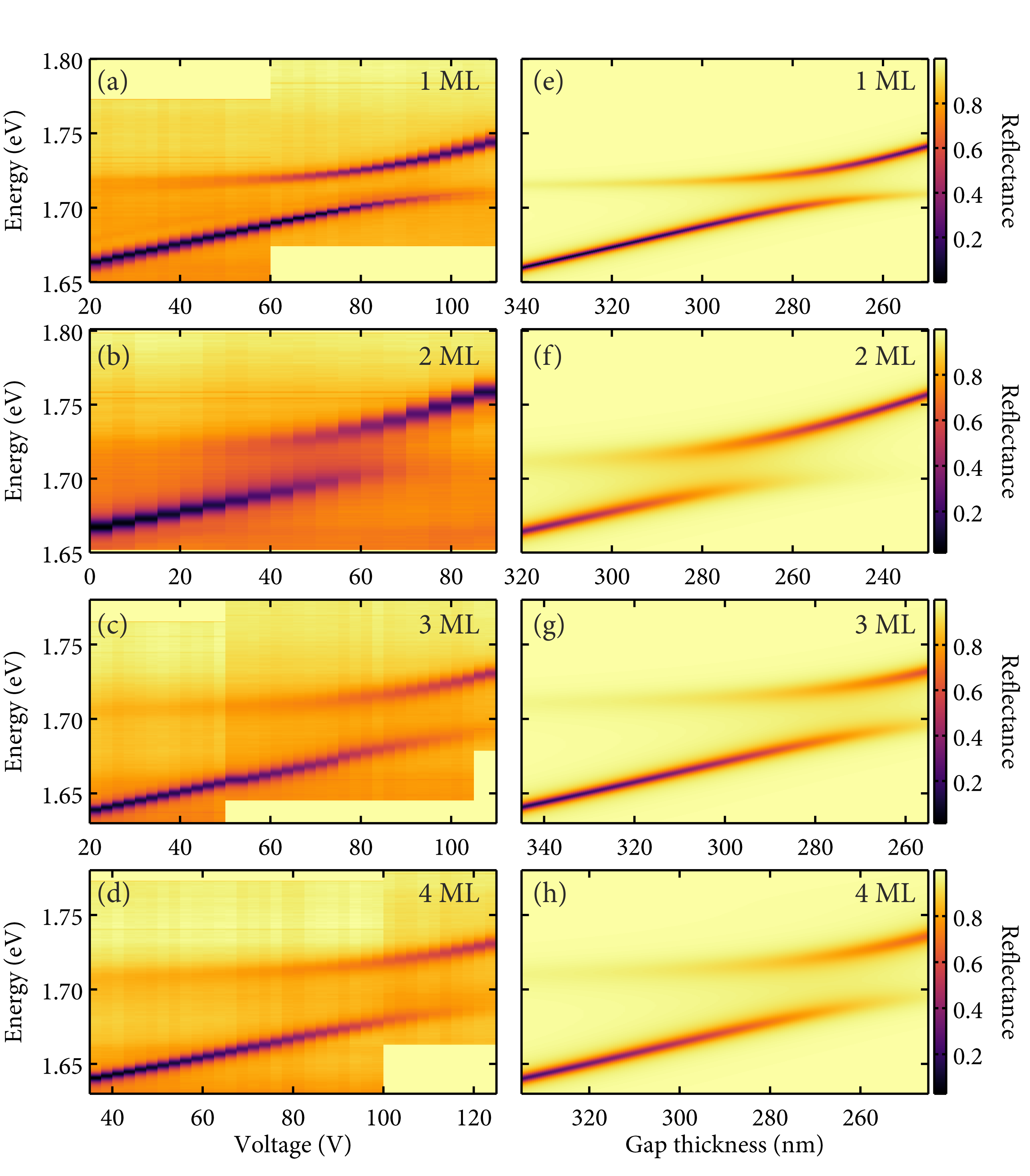}
\caption{Left column: maps showing experimental reflectance spectra for different voltages applied to a piezoelement tuning the cavity width, measured at 5\,K and normal incidence on hBN-encapsulated: (a)~1\,ML, (b)~2\,ML, (c)~3\,ML, (d)~4\,ML WSe\textsubscript{2} flakes. Right column: transfer matrix method simulations of the experimental data shown in the left column, performed for: (e)~1\,ML, (f)~2\,ML, (g)~3\,ML, (h)~4\,ML as a function of distance between the DBR mirrors.}
\label{im:FigCoupling}
\end{figure*}

We investigated 1\,ML- to 4\,ML-thick WSe\textsubscript{2} flakes  embedded in an open, tunable dielectric cavity schematically shown in Fig.\,\ref{im:Fig1}(a). The cavity consists of two dielectric distributed Bragg reflectors (DBRs), each made of 5 pairs of SiO\textsubscript{2}/TiO\textsubscript{2} layers with the maximum reflectance at 723\,nm (1.715\,eV), chosen specifically to match the energy of the A exciton in monolayer WSe\textsubscript{2} encapsulated in hBN \cite{Manca_NatComm2017}. The mirrors were  grown on a transparent fused silica substrate to provide  optical access also from the  back side.

WSe\textsubscript{2} flakes were mechanically exfoliated from commercially available, nominally undoped bulk crystals. Selected flakes of 1 to 4\,ML thickness were then transferred onto approx.\,183\,nm thick (see Fig.\,S1 Supplementary Materials) hBN flakes previously deposited on top of the bottom DBR. The thickness of the bottom hBN flakes was chosen to obtain the local maximum of the electric field standing wave inside the final full cavity at the position of the WSe\textsubscript{2} flakes to assure the efficient exciton-photon coupling conditions (see Supplementary Materials for details). The hBN capping layer covering the WSe\textsubscript{2} flakes was much thinner, approx.\,10\,nm. Fig.\,\ref{im:Fig1}(b) presents an optical microscope image of the investigated WSe\textsubscript{2} flakes. The hBN encapsulation allows to significantly reduce the inhomogeneous broadening of spectral lines, leading to linewidths that approach the radiative decay limit \cite{Cadiz_PRX2017,Ajayi_2DMater2017}. 

The influence of the encapsulation is still under investigation \cite{Fang_arxiv2019}, but hBN provides isolation of the electrostatic disorder at the SiO\textsubscript{2} interfaces and acts also as an atomically flat substrate for the flakes. A comparison of surface roughness of the DBR and the hBN flake deposited on that DBR, measured by atomic force microscopy, is shown in the inset of Fig.\,\ref{im:Fig1}(c).

The full cavity, presented in Fig.\,\ref{im:Fig1}(a), is formed by covering the structure discussed above with another DBR. We performed angle-resolved reflectance measurements at 5\,K revealing the strong coupling regime between the cavity photons and the excitons in all 1- to 4\,ML-thick WSe\textsubscript{2}  flakes, as presented in Fig.\,\ref{im:Fig1}(d--f). In an angle-resolved experiment, the cavity photon energy (Ph) depends quadratically on the in-plane wavevector, which is proportional to the emission angle. The strong coupling regime of the cavity mode with the exciton resonances in thin WSe\textsubscript{2} layers (X$_i$)  manifests itself by the appearance of two anticrossing lower and upper polariton branches (LP and UP), with dispersion relations well described by a two coupled oscillators model, as demonstrated with dashed white  curves in Fig.\,\ref{im:Fig1}(d--f).

\begin{figure}
\centering
\includegraphics[width=7.4cm]{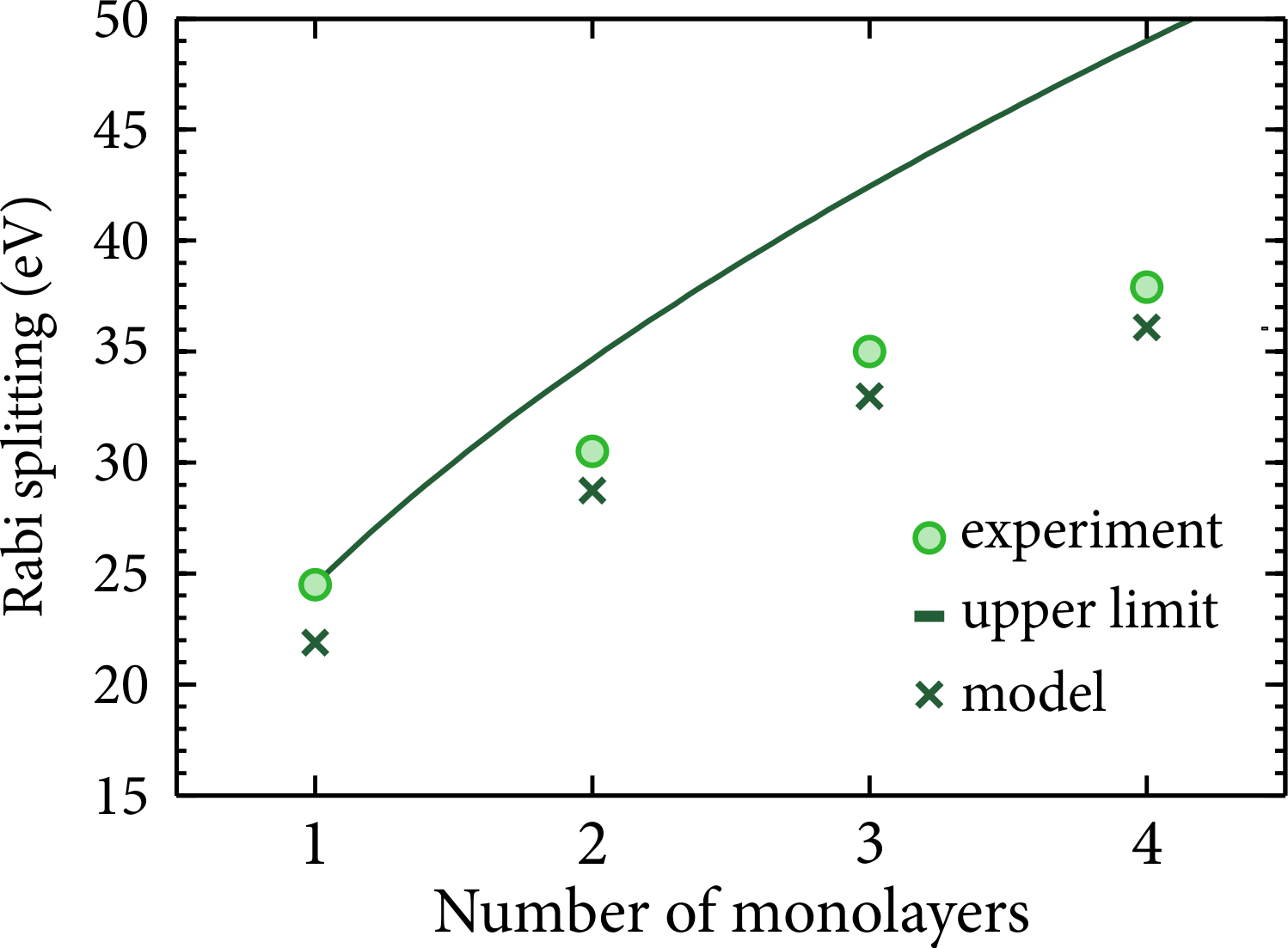}
\caption{Coupling strength between the photonic cavity mode and the excitonic resonance in multilayer WSe\textsubscript{2}. The upper limit shows the highest possible coupling strength that could be achieved for a structure consisting of multiples of WSe\textsubscript{2} monolayers. Marked with crosses are modelled values of the multilayer coupling strength obtained from calculations based on a transfer matrix method.}
\label{im:FigOmega}
\end{figure}

The energy of photons confined in the cavity was tuned in a continuous way by smoothly changing the distance between the mirrors \cite{Krol_Nanoscale2019}. Experimentally, it was realized by placing one of the mirrors on a piezoelectric stage and applying to it an external bias. Reflectance spectra at normal incidence for variable electrical bias applied to the piezoelectric chip taken at the position of a specific WSe\textsubscript{2} flake are presented in Fig.\,\ref{im:FigCoupling}.  Upon applying the voltage, the distance between the DBRs decreases, which results in the increase of the cavity photon energy. When the mode approaches the energy of the excitonic resonance in a given WSe\textsubscript{2} flake, a clear anticrossing behaviour can be observed. The minimal energy separation between the modes is equal to the coupling strength $\Omega$, also referred to as the vacuum field Rabi splitting. The observed coupling strength increases with the number of  WSe\textsubscript{2} layers, as shown in Fig.\,\ref{im:FigOmega}.

The coupling strength between a photonic mode of a planar cavity incorporating an optically active layer with an excitonic resonances can be in general well approximated by \cite{Flatten_SciRep2016,Savona_1995}:
\begin{equation}
\Omega = 2 \sqrt{\frac{f d}{L n_c^2}},
\label{eq:omega}
\end{equation}
where $L$ is the effective cavity length, $n_c$ is the cavity refractive index and $f$ is proportional to the exciton oscillator strength in a layer of thickness $d$. The coupling strength given by eq.\,(\ref{eq:omega}) can provide the upper limit for the Rabi splitting: one can expect an increase of the coupling strength proportional to the square root of the active layer thickness, as plotted with a solid line in Fig.\,\ref{im:FigOmega}. In our experiment we observed a slower increase of the coupling strength with the number of WSe\textsubscript{2} layers  as compared to that given by  the upper limit described above and derived for the monolayer case. Such behaviour can be caused by two factors: change of the exciton resonance width or decrease of the exciton oscillator strength  in multilayer WSe\textsubscript{2}  with respect to the monolayer.

\begin{figure*}
\centering
\includegraphics[width=\textwidth]{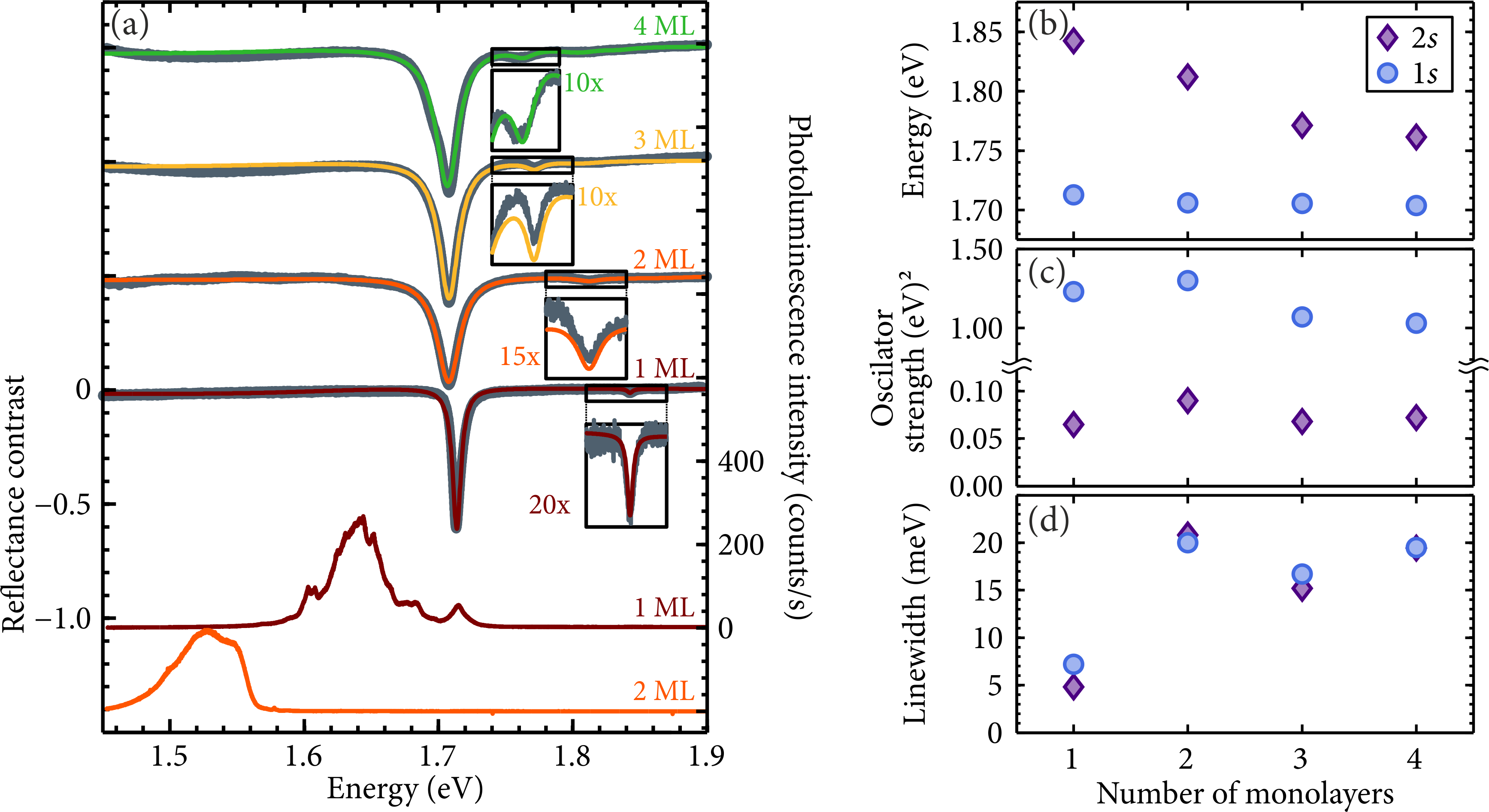}
\caption{(a) Reflectance contrast (RC) and photoluminescence (PL) spectra of hBN encapsulated WSe\textsubscript{2} multilayers measured without the top DBR at 5\,K. Experimental RC spectra are marked with grey lines, whereas the color  curves represent the results of fitting the data with a transfer-matrix-based model. Shown in the right column are the fitting parameters of the ground and first excited exciton resonances in multilayer WSe\textsubscript{2}:  the energy position (b),  the oscillator strength (c) and  the linewidth (d).}
\label{im:FigRC}
\end{figure*}

\begin{table*}
\caption{RC fitting parameters for hBN encapsulated WSe\textsubscript{2} layers deposited on  a DBR  and a comparison of the modelled exciton-photon coupling strength in a dielectric cavity (abbrev. model)  to the values obtained experimentally (abbrev. exp.).}
\label{TabhBN}
\begin{tabular}{lcccccccccc}
\textrm{} &	\multicolumn{1}{c}{\textrm{$E_{1s}$ (eV)}} &	\multicolumn{1}{c}{\textrm{$f_{1s}$ (eV\textsuperscript{2})}}	& \multicolumn{1}{c}{\textrm{$\gamma_{1s}$ (meV)}}& \hspace*{0.05cm} &  \multicolumn{1}{c}{\textrm{$\Omega_{1s}^{\rm model}$ (meV)}} & \multicolumn{1}{c}{\textrm{$\Omega_{1s}^{\rm exp}$ (meV)}} & \hspace*{0.05cm} &	\multicolumn{1}{c}{\textrm{$E_{2s}$ (eV)}} &	\multicolumn{1}{c}{\textrm{$f_{2s}$ (eV\textsuperscript{2})}}	& \multicolumn{1}{c}{\textrm{$\gamma_{2s}$ (meV)}} \\

\hline

\textrm{1 ML}   & 1.7128 & 1.23  & ~7.2 & & 21.9 & 24.5 & & 1.8426 & 0.065 & ~4.8\\
\textrm{2 ML}   & 1.7059 & 1.30  & 20.0 & & 28.7 & 30.5 & & 1.8122 & 0.090 & 20.8\\
\textrm{3 ML}   & 1.7055 & 1.07  & 16.7 & & 33.0 & 35.0 & & 1.7713 & 0.068 & 15.2 \\
\textrm{4 ML}   & 1.7036 & 1.03 & 19.5 & & 36.9 & 36.1 & & 1.7632 & 0.072 & 19.4\\

\end{tabular}
\end{table*}

To evaluate the influence of both  these factors we performed reflectance contrast (RC) and photoluminescence (PL) measurements without the top DBR. RC spectra  taken  on each WSe\textsubscript{2} flake  are presented in Fig.\,\ref{im:FigRC}. The spectra are dominated by a pronounced dip around 1.71\,eV, corresponding to the absorption of a neutral free exciton formed at the $K^\pm$ valleys, i.e. the so-called A exciton \cite{Koperski_Nanophotonics2017}. 
For a monolayer, this excitonic transition is also observable in the PL spectrum, which is dominated by a broad band attributed to the so-called localised excitons \cite{Arora_Nanoscale2015,Smolenski_PRX2016,Smolenski_2017,Koperski_2018}. 
With the increase of WSe\textsubscript{2} layer thickness, the energy of the excitonic resonance observed in the RC slightly decreases and broadens, with the most significant difference occurring between the monolayer and the bilayer. For WSe\textsubscript{2} thicker than the monolayer, as the bandgap changes from direct to indirect, the transitions related to the A excitons  can no longer be observed in the PL spectra. For 2\,MLs, the  emission occurring at around 1.53\,eV is related to the indirect bandgap transitions \cite{Arora_Nanoscale2015}.
 
 The high optical quality of hBN encapsulated flakes allow to observe in RC the first excited exciton states, i.e. 2\textit{s} \cite{Chernikov_PRL2014,He_PRL2014,Chen_PRL2018,Molas_2019}, shown in the insets in Fig.\,\ref{im:FigRC}(a). Their energy separation from the main absorption line significantly decreases with  the number of layers due to the expected decrease of the neutral exciton's binding energy \cite{Chernikov_PRL2014}.

To extract both the exciton oscillator strength and exciton resonance linewidth, the RC spectra were fitted using a transfer matrix method. The RC signal was calculated based on simulated reflectance of the DBR and the DBR with a corresponding WSe\textsubscript{2} layer encapsulated in hBN. The exciton resonance was introduced as a dielectric function of the WSe\textsubscript{2} layer given by a Lorentz model in  the form  provided in \cite{Li_PRB2014}: 
\begin{equation}
\varepsilon_{\mathrm{WSe\textsubscript{2}}}\!\left(E\right) = \varepsilon_0 - \sum_i\frac{f_{i}}{E_{i}^2-E^2 - {\rm i} \gamma_{i} E},
\label{eq:lorentz}
\end{equation} 
where $i$ describes the excitonic ground state and the first excited 2$s$ state. The optical response of the system is also directly dependent on the thicknesses inserted into the transfer matrix model, which makes $f_i$ defined in eq. (\ref{eq:lorentz}) a measure of the transition strength per unit thickness of a given WSe\textsubscript{2} layer. Depending on  the number of constituent monolayers, WSe\textsubscript{2} the thickness of a given WSe\textsubscript{2} flake was taken as a multiple of the  single layer thickness equal to 0.65\,nm \cite{Radisavljevic_NatNanotech2011}.
The dielectric constant away from the resonances was taken as $\varepsilon_0 = 16$ \cite{Li_PRB2014}.

The transfer matrix model  curves were directly fitted  to the experimental RC spectra with $f_{i}$, $E_{i}$ and $\gamma_{i}$ treated as fitting parameters.
 The results are presented  with solid color lines in Fig.\,\ref{im:FigRC}. All fitting parameters are also summarised in Tab.\,\ref{TabhBN}. Fig.\,\ref{im:FigRC}(b)  shows the energies of excitonic resonances \ in hBN encapsulated WSe\textsubscript{2} flakes of different thicknesses. The ground state energy is almost independent of the layer thickness, whereas the energy of the excited exciton state significantly decreases. For the monolayer, the 2\textit{s} state is 130\,meV above the ground state and for the tetra-layer this energy difference decreases to 61\,meV. Also the exciton's ground state oscillator strength [see Fig.\,\ref{im:FigRC}(c)]  slightly decreases with  the number of layers. For the excited exciton states the oscillator strength is 19 to 14 times smaller than for the corresponding ground state with no evident dependence on the number of layers.

    It is worth to mention that due to  low oscillator strength of the excited 2\textit{s} exciton state in our samples,  we do not observe a strong coupling regime at this transition, as shown in Fig.\,S9. Such a coupling with excited exciton states  has recently become of interest due to large optical nonlinearities one can probe in this type of systems \cite{Gu2019,Walther_NatCommun2018}. In our structures, the 2\textit{s} exciton-polariton is expected to occur with $\Omega=3.2$\,meV  (see Fig.\,S10), which is much lower than the linewidth of the 2\textit{s} resonance as shown in Fig.\,\ref{im:FigRC}(c).

The exciton resonance linewidth determines the optical quality of the layers and is usually affected by homogeneous and inhomogeneous contributions. For both the ground state and the excited 2$s$ state, the spectrally narrowest transition occurs in the monolayer. For the bilayer, the linewidth increases over two times, and then is almost constant for 3- and 4\,ML-thick layers. Starting from the 3\,ML thick flake, the excitonic resonances reveal a double structure. The appearance of such a fine structure (an increase of the linewidth or a presence of  two partially overlapping resonances) can be explained in terms of  hybridisation occurring predominantly in the valence band of thicker TMD layers (>1\,ML), due to different symmetries of the conduction and valence band orbitals (see Ref. \cite{molasNanoscale} for details). The analogous effect has already been observed for multilayers of WS\textsubscript{2} deposited on Si/SiO\textsubscript{2} substrate \cite{molasNanoscale} and  for mono- and few-layer MoS\textsubscript{2} encapsulated in hBN \cite{Slobodeniuk_2DMater2019,Gerber_PRB2019}.

The exciton properties obtained from the RC measurements of the WSe\textsubscript{2} layers deposited on the DBRs were used to simulate their behaviour inside a full cavity structure, as presented in Fig.\,\ref{im:FigCoupling}(e--h). We used the transfer matrix  method to simulate the reflectance at normal incidence from a whole structure consisting of two DBRs and a WSe\textsubscript{2} flake encapsulated in hBN for variable spacing between the mirrors. The calculations well reproduce the experimental reflectance maps from the left column of Fig.\,\ref{im:FigCoupling}. The exciton--photon coupling strength was extracted from the reflectance maps as  the minimum energy distance between the two polaritonic branches, and  then included in Fig.\,\ref{im:FigOmega}. As can be seen, the  simulations slightly underestimate the coupling strength, but properly reconstruct the trend of increasing coupling strength with the thickness of WSe\textsubscript{2} layer.


\section{Summary}

In summary, we have demonstrated a strong coupling regime between photons in a planar optical cavity  and excitons in multilayer WSe\textsubscript{2}. We have investigated flakes encapsulated in hBN as well as unprotected ones. The strong coupling is observed regardless of the crossover from direct to indirect bandgap  accompanying the transition from  the monolayer to  $N$-layer thickness. Despite the decrease of the excitons' oscillator strength per layer thickness at the $K\pm$ points of the Brillouin zone in few-layer WSe\textsubscript{2}, the observed coupling strength increases with the thickness of an active WSe\textsubscript{2} layer  which proves that multilayers can be well adopted to operate in the strong light-matter coupling regime with substantial absorption at the polariton modes.  
By changing the number of WSe\textsubscript{2} layers and controlling the distance between the Bragg mirrors we were able to tune the energies of  the system eigenstates -- the exciton-polaritons. Recently, it  has been demonstrated that exciton-polaritons in TMDs can be coupled to the emission of a semiconductor quantum well \cite{Wurdack_NatComm2017,Waldherr_NatCommun2018}, or organic dye \cite{Flatten_NatComm2017} forming hybrid TMDs-semiconductor-light states. Although indirect transitions  analyzed in this paper would decrease the efficiency of such emission, the TMD multilayers should definitely be considered as potential building blocks for future  engineering of hybrid TMDs eigenstates.

\section*{Methods}

\subsection*{Optical measurements}

All optical measurements were performed  on samples placed in a cryostat at the temperature of 5\,K. 
PL signal was excited nonresonantly with a 514.5\,nm (2.41\,eV) laser beam. The excitation power, measured before the entrance of a microscope objective, was equal to 100\,$\upmu$W and  concentrated in approx. 1\,$\upmu$m spot. For reflectance and RC measurements a broadband halogen lamp was used with a spot size of approx. 5\,$\upmu$m. 

Investigations of the cavity structures were performed with angular resolution, achieved by imaging the Fourier space of a high numerical aperture ($\mathit{NA} =0.55$) microscope objective. Data shown in Fig.\,\ref{im:FigCoupling} corresponds to reflectance at normal incidence extracted from the angle-resolved spectra. 

Demonstration of strong light-matter coupling regime and formation of the exciton-polariton modes  in 1 to 6\,ML thick WSe\textsubscript{2} flakes, which  have not been encapsulated in hBN, can be found in supplementary material.

\section*{Acknowledgements}

The authors thank Artur Slobodeniuk for helpful discussions. This work has been supported by the Ministry of Higher Education Republic of Poland budget for education via the research projects "Diamentowy Grant" 0109/DIA/2015/44 and 0005/DIA/2016/45; by the National Science Centre: grants 2013/10/M/ST3/00791, 2018/31/B/ST3/02111 and 2018/31/N/ST3/03046; by the EU Graphene Flagship project (ID: 785219), and the ATOMOPTO project (TEAM programme of the Foundation for Polish Science co-financed by the EU within the ERD Fund). B.P. acknowledges Ambassade de France en Pologne for the research stay project. K.W. and T.T. acknowledge support from the Elemental Strategy Initiative conducted by the MEXT, Japan and the CREST (JPMJCR15F3), JST.



%

\end{document}


\title{Supplementary information for Exciton-polaritons in multilayer WSe\textsubscript{2} in a planar microcavity}

\author{M.\,Kr\'ol}
\author{K.\,\,Rechci\'nska}
\author{K.\,Nogajewski}
\author{M.\,Grzeszczyk}
\author{K.\,\L{}empicka}
\author{R.\,Mirek}
\author{S.\,Piotrowska}
\affiliation{Institute of Experimental Physics, Faculty of Physics, University of Warsaw, ul.~Pasteura 5, PL-02-093 Warsaw, Poland}
\author{K.\,Watanabe}
\author{T.\,Taniguchi}
\affiliation{National Institute for Materials Science, Tsukuba, Ibaraki, 305-0044, Japan}
\author{M.\,R.\,Molas}
\affiliation{Institute of Experimental Physics, Faculty of Physics, University of Warsaw, ul.~Pasteura 5, PL-02-093 Warsaw, Poland}
\author{M.\,Potemski}
\affiliation{Institute of Experimental Physics, Faculty of Physics, University of Warsaw, ul.~Pasteura 5, PL-02-093 Warsaw, Poland}
\affiliation{Laboratoire National des Champs Magn\'etiques Intenses, CNRS-UGA-UPS-INSA-EMFL, Grenoble, France}
\author{J.\,Szczytko}
\author{B.\,Pi\k{e}tka}
\email{Barbara.Pietka@fuw.edu.pl}
\affiliation{Institute of Experimental Physics, Faculty of Physics, University of Warsaw, ul.~Pasteura 5, PL-02-093 Warsaw, Poland}

\begin{abstract}
\end{abstract}

\maketitle

\section{\lowercase{h}BN characterisation }

To ensure the most effective light--matter coupling the \ch{WSe2} layer should be positioned at the antinode of the standing wave of the electromagnetic field inside the cavity. For cavity without hBN such requirement can be fulfilled when DBRs are ending with lower refractive index material: \ch{SiO2}. Using the same DBRs for encapsulated \ch{WSe2}, lower hBN layer should have optical thickness of $\lambda_0/2$.  For the excitonic transition in monolayer \ch{WSe2} at approx. 710\,nm and hBN refractive index of 2.1 \cite{Lee_PSSB2018}, this corresponds to approx. 170\,nm. 

\begin{figure*}
\centering
\includegraphics[width=\textwidth]{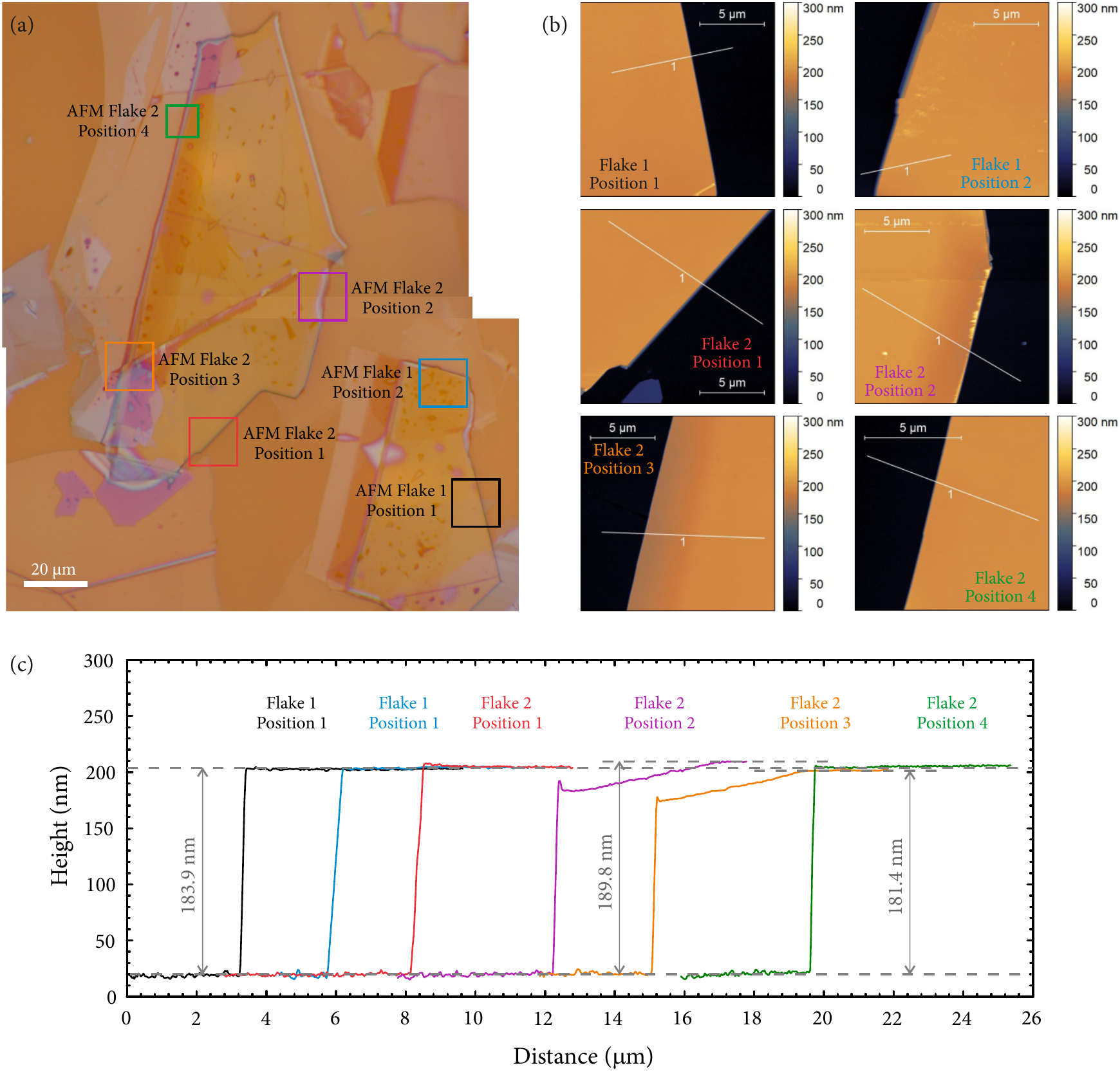}
\caption{Characterization of hBN layers. (a) Microscopic image of DBR surface with deposited exfoliated hBN. Marked regions were investigated using AFM and are presented in (b).  (c) Height profiles along straight lines marked in (b).  }
\label{im:FighBN}
\end{figure*}

The exfoliated hBN flakes with comparative thickness were transferred to the DBRs. Images of the transferred hBN with the corresponding height profiles measured by AFM are presented in \figref{im:FighBN}.

\section{Empty cavity}

\begin{figure*}
\centering
\includegraphics[width=\textwidth]{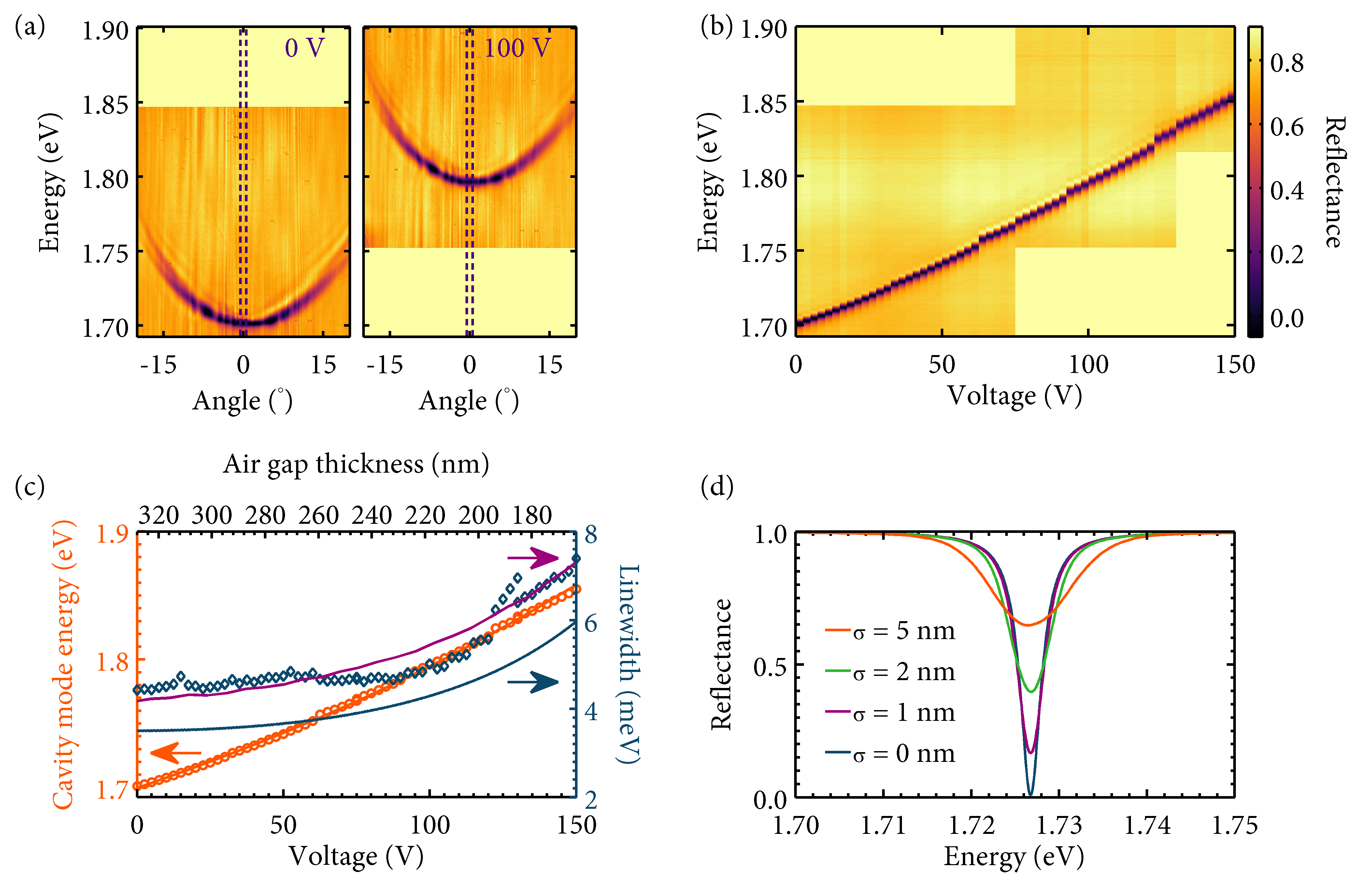}
\caption{Tunable cavity with with 183\,nm thick hBN layer. (a) Angle-resolved reflectance spectra at two different voltages applied to the piezoelectric actuator.  (b) Tuability of the cavity at normal incidence [range between the dashed lines in (a)].  (c) Experimental values of cavity mode energy and reflection deep linewidth change with bias applied to the piezoelectric chip (points) compared with transfer matrix simulations for decreasing air gap thickness (lines). Blue line shows linewidth for a perfectly parallel mirrors. Purple line marks simulation with assumption that the distance between the mirror is a random variable described by normal distribution with a dispersion $\sigma = 1$\,nm. (d) Simulations of normal incidence reflectance from the cavity where the distace between the mirrors is given by normal distribution with mean value of 300 and given dispersion $\sigma$.   }\label{im:FigCavity}
\end{figure*}

Reflectance from a tunable cavity without active \ch{WSe2} flakes is presented in \figref{im:FigCavity}. Measurements were taken from the position of 183\,nm thick hBN layer. All of the measurements of the cavity structures were angle-resolved. \figref[(a)]{im:FigCavity} shows reflectance maps obtained at two different electric biases applied to the piezoelectric chip, which was used to tune the distance between the mirror. Full tunability range for voltages 0--150\,V is shown in \figref[(a)]{im:FigCavity}, which presents reflectance at normal incidence. Data for incidence perpendicular to the cavity plane was selected from the angle-resolved maps.  The cavity mode resonance energy shifts over 150\,meV linearly with voltage. 

In \figref[(c)]{im:FigCavity} experimental position and linewidth of the cavity mode is compared with transfer matrix simulations. The air-gap thickness taken for simulations were extracted from the experimental energy position of the cavity mode.  Reflectance deep linewidth, at the order of 5\,meV, is almost constant up to 110\,V, and slightly increases for thinner cavity. This behaviour is consistent with transfer matrix simulations marked by blue line in \figref[(c)]{im:FigCavity}, but the calculated linewidths are slightly lower than is observed experimentally. Such broadening of experimental results can be explained when the mirrors forming the cavity are not perfectly parallel. Measuring reflectance with a 5\,$\upmu$m Gaussian spot we average over such area. This averaging can be taken into account in transfer matrix simulations with assumption, that the distance between the mirrors doesn't have a finite value, but rather is a random variable. Calculations performed for the intermirror distance described by a normal distribution with a small dispersion $\sigma$ equal to 1\,nm, shows better agreement between linewidth observed in the experiment and the model [purple line in \figref[(c)]{im:FigCavity}].  
   
\figref[(d)]{im:FigCavity}] presents reflectance spectra given by such modified for broadening model for different values of the dispersion of the normal distribution.

\section{WS\lowercase{e}\textsubscript{2} Flakes without \lowercase{h}BN encapsulation}

As a reference we prepared another sample with WSe\textsubscript{2} flakes without hBN encapsulation. Those 1 to 6\,ML thick flakes were deposited directly onto a surface of a identical DBR, as described in the main text. Performed reflectance contrast (RC) and photoluminescence (PL) mesurements of the on top mirror are presented in \figref[(a)]{im:nohBNRC}. 

For the WSe\textsubscript{2} monolayer the RC spectrum measured in the vicinity of the optical bandgap,  consists of a single narrow deep at 1.753\,eV,  corresponding to the absorption of a neutral free exciton (so-called A exciton) \cite{Koperski_Nanophotonics2017}. For the bilayer, the absorption transition is observed at lower energies and significantly broadens. Starting from the 3\,ML thick flakes, the related resonaces reveal a doubled structure. Appearance of such fine structure can be explained by the hybridisation of the valence and conduction bands in TMD layers (>1\,ML), as has already been observed for multilayers of WS\textsubscript{2} deposited on Si/SiO\textsubscript{2} substrate \cite{molasNanoscale} and of MoS\textsubscript{2} encapsulated in hBN \cite{Slobodeniuk_2DMater2019,Gerber_PRB2019}

For monolayer, the PL spectrum is dominated by a broad band attributed to the so-called localised excitons \cite{Arora_Nanoscale2015,Smolenski_PRX2016,Koperski_2018}. The inset in \figref[(a)]{im:nohBNRC} shows a weaker peak at higher energies, which is associated with the emission of free neutral A excitons. Its energy coincides with the energy of the corresponding A resonance apparent in the RC spectrum. The emission energy and intensity decrease significantly for thicker WSe\textsubscript{2} flakes, as the bandgap change from direct one for the ML to indirect one in multilayers. For the 2 and 3\,MLs, the observed emission occurring correspondingly at around 1.55\,eV and 1.42\,eV is related to the indirect band gap \cite{Arora_Nanoscale2015}.

\section{Flakes deposited on \lowercase{h}BN without capping layer}

We investigated also two additional monolayer and a bilayer \ch{WSe2} flakes isolated from the DBR by hBN, but without the thin covering hBN flake. Photoluminescence and reflectance contrast spectra without the top mirror are shown in \figref{im:Fig1BN} for monolayer and in \figref{im:Fig2BN} for  bilayer. Optical properties of those flakes, like linewidth and emission intensity,  are similar to the fully encapsulated \ch{WSe2}. The excitonic transition energies are in between dielectric-screened encapsulated flakes and the deposited directly on the DBR. 2\textit{s} excited exciton transitions are not observed in the reflectance measurement.

\section{Coupling with 2\textit{\lowercase{s}} exciton state}

Tuning the cavity mode to the 2\textit{s} exciton resonance energy visible on reflectance contrast spectra for encapsulated monolayer flake does not reveal any sign of coupling due to low value of oscillator strength of this transition. For the cavity mode energy crossing the 2\textit{s} exciton resonance energy no sign of anticrossing or mode linewidth broadening is observed, as shown in \figref{im:Fig2s}. No sign of coupling can be seen also at angle resolved spectra shown in  \figref[(a,b)]{im:Fig2s}.

Such observation is consistent with the transfer matrix simulations. \figref[(a)]{im:Fig2sModel} shows calculated reflectance spectra for zero detuning between the cavity mode and a 2\textit{s} exciton state for different values of oscillator strength. For the value determined by the fitting to the reflectance contrast measurement, simulated  
 reflectance spectra can be well described by a sum of two Lorentz functions, but with the distance between them lower than their widths.

\begin{figure}[!htb]
\centering
\includegraphics[width=.495\textwidth]{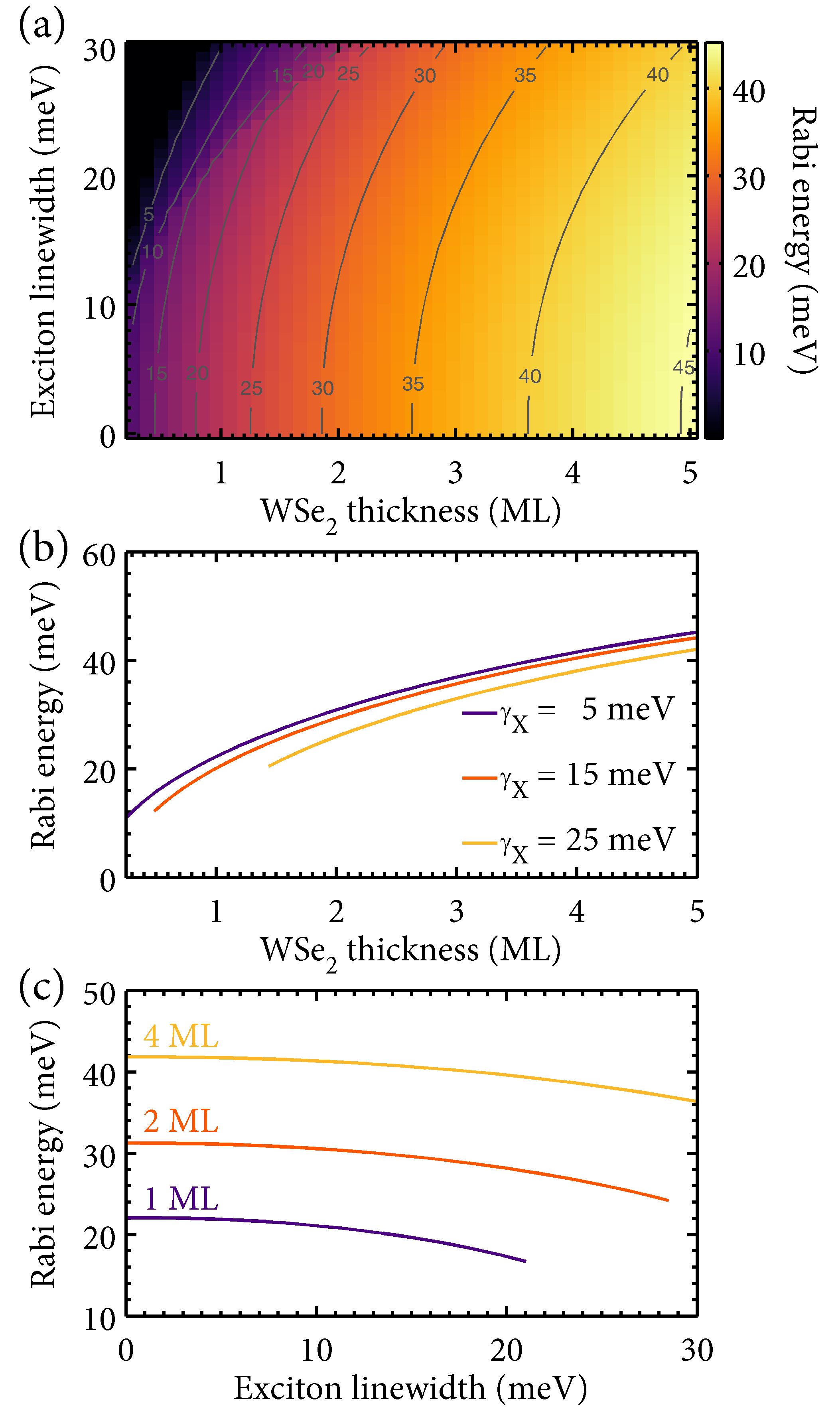}
\caption{Transfer matrix simulations for WSe\textsubscript{2} layer with \mbox{$f = 1.23$} and encapsulated in hBN. (a) Rabi energy dependence on (b) WSe\textsubscript{2} layer thickness (c) exciton resonance linewidth.}
\label{im:dFWHMomega}
\end{figure}

\begin{figure*}[!htb]
\begin{minipage}[t]{\textwidth}
\centering
\includegraphics[width=\textwidth]{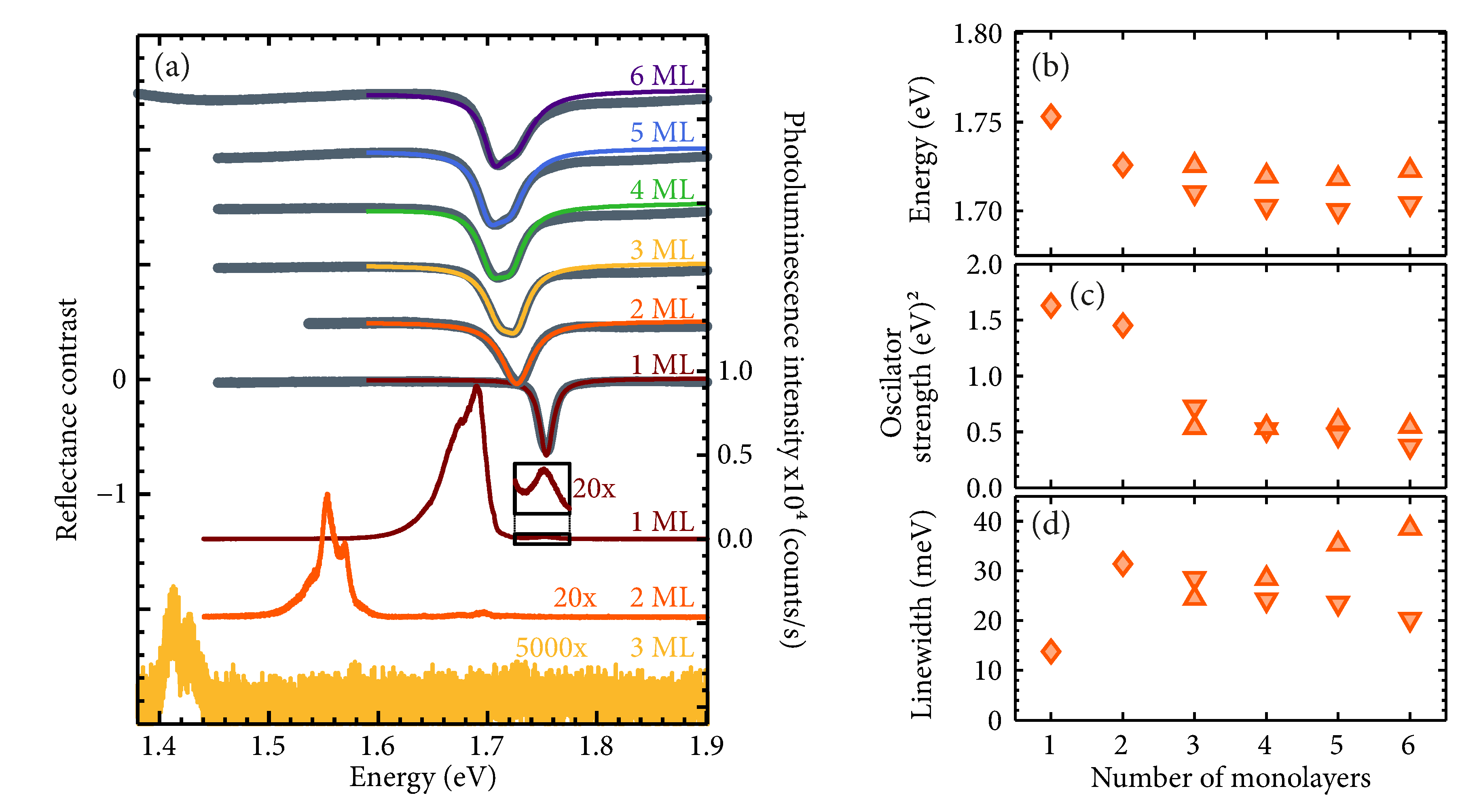}
\caption{(a) Reflectance contrast (RC) and photoluminescence (PL) spectra of WSe\textsubscript{2} multilayers depositedon top of a DBR measured at 5\,K. Experimental RC spectra are marked with gray lines, whereas colour lines represent the fitted transfer matrix model. Fitting parameters of ground and first excited exciton resonances in multilayer WSe\textsubscript{2}: (b) energy positions, (c) oscillator strength and (d) linewidth. $\bigtriangleup$ mark the fine structure components of higher energy and $\bigtriangledown$ of the lower one.}
\label{im:nohBNRC}
\end{minipage}


\begin{minipage}[t]{\textwidth}
\captionof{table}{RC fitting parameters for WSe\textsubscript{2} deposited on the DBR with modeled  exciton-photon coupling strength in dielectric cavity (abbrev. model) compared with the values obtained experimentally (abbrev. exp).}
\label{TabNohBN}
\begin{ruledtabular}
\begin{tabular}{lcccccc}
\textrm{} &	\multicolumn{1}{c}{\textrm{$E_{1s}$ (eV)}} &	\multicolumn{1}{c}{\textrm{$f_{1s}$ (eV\textsuperscript{2})}}	& \multicolumn{1}{c}{\textrm{$\gamma_{1s}$ (meV)}} &	 \hspace*{0.05cm} &  \multicolumn{1}{c}{\textrm{$\Omega_{1s}^{\rm model}$ (meV)}} & \multicolumn{1}{c}{\textrm{$\Omega_{1s}^{\rm exp}$ (meV)}}\\
\hline
				\textrm{1 ML}   & 1.7532 &  1.63 & 13.8 & ~ & 19.4 & 23.9 \\
				\textrm{2 ML}   & 1.7258 &  1.45 & 31.4 & ~ & 20.8 & 23.2 \\
				\textrm{3 ML}   & 1.7103 &  0.72 & 28.5 & ~ & 29.1 & 30.2 \\
							    & 1.7257 &  0.54 & 24.5 & ~ &    ~ & \\
 				\textrm{4 ML}   & 1.7026 &  0.52 & 24.2 & ~ & 31.3 & 37.5\\
 				                & 1.7196 &  0.54 & 28.5  \\
				\textrm{5 ML}   & 1.7002 &  0.47 & 23.5 & ~ & 33.6 & 36.8\\
				                & 1.7182 &  0.59 & 35.4 \\
 				\textrm{6 ML}   & 1.7041 &  0.37 & 20.3 & ~ & 33.2 & 40.7\\
 				                & 1.7228 &  0.55 & 38.6 \\
\end{tabular}
\end{ruledtabular}
\end{minipage}

\includegraphics[width=.45\textwidth]{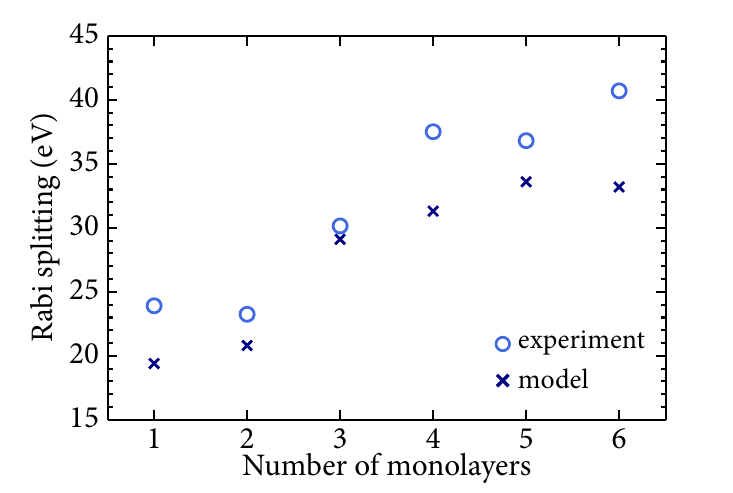}
\caption{Experimental coupling strength of multilayer WSe\textsubscript{2} flakes.  Modelled values of coupling strength based on reflectance spectra calculated with transfer matrix method are marked with crosses.}
\end{figure*}

\begin{figure*}[!htb]
\centering
\includegraphics[width=.45\textwidth]{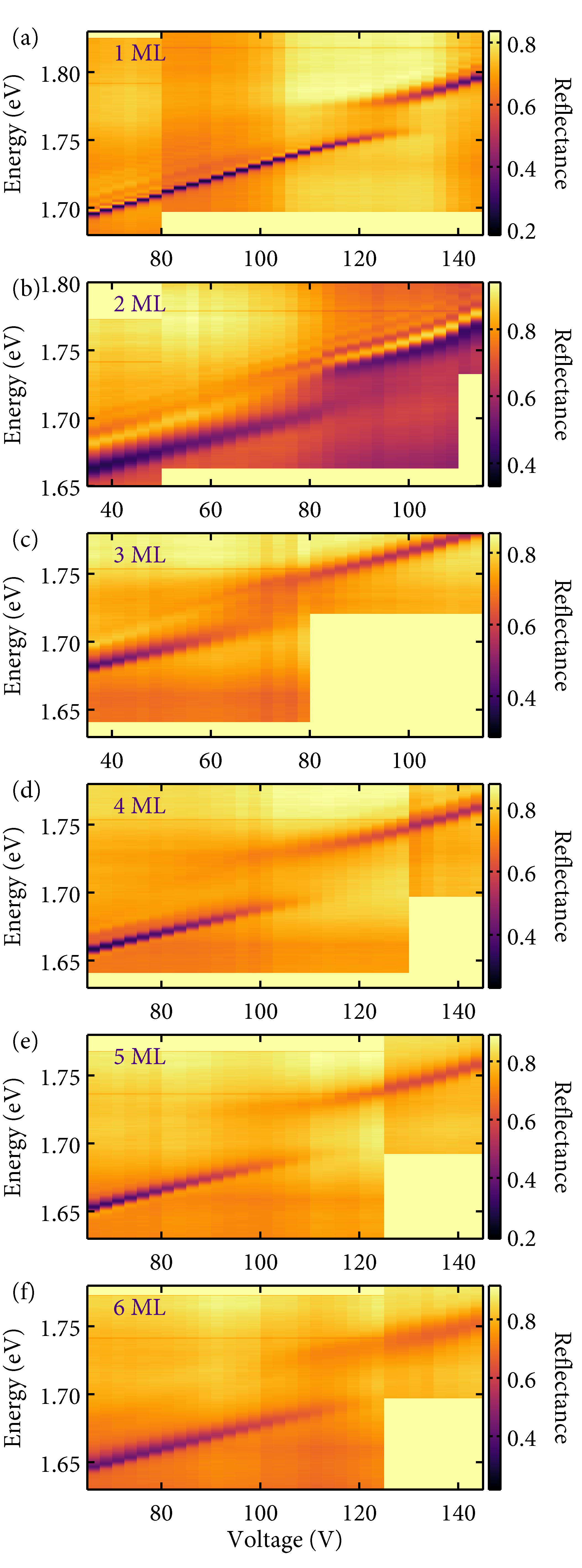}\quad
\includegraphics[width=.45\textwidth]{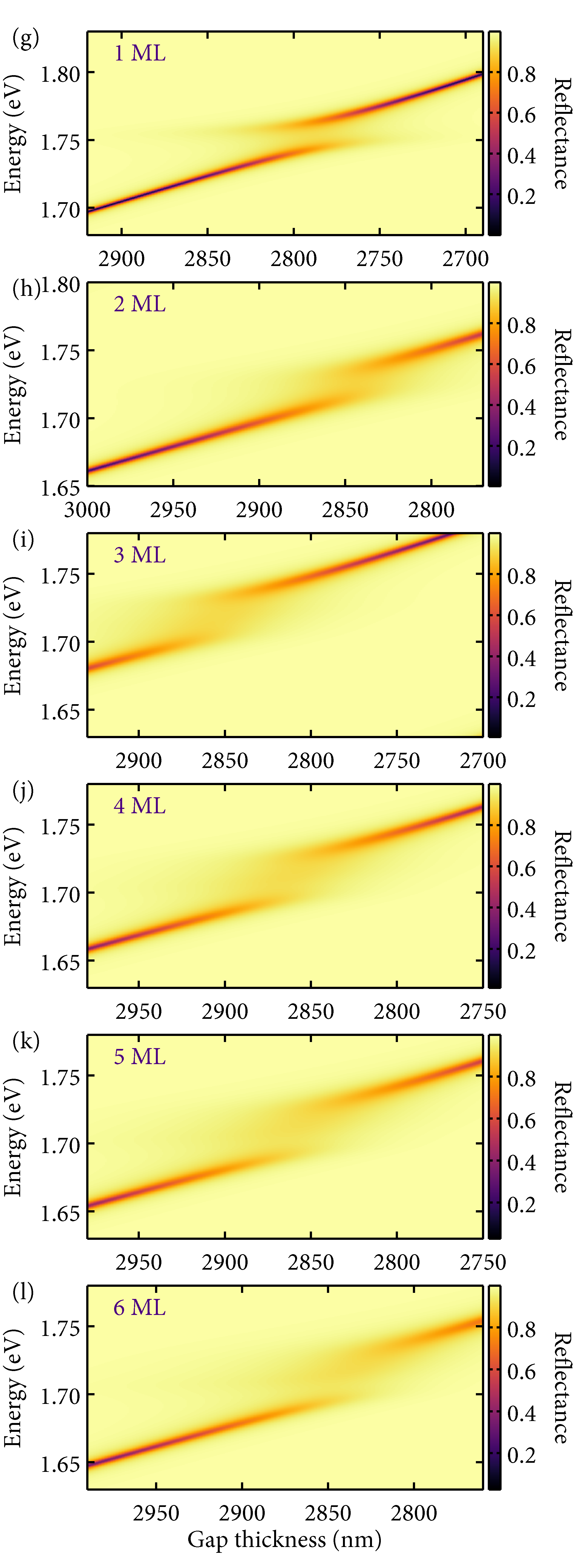}
\caption{\ch{WSe2} few-layer flakes in tunable cavity. Reflectance map for normal incidence for tuning the cavity with voltage applied to the piezoelement with: (a)~1\,ML, (b)~2\,ML, (c)~3\,ML, (d)~4\,ML, (e)~5\,ML and (f)~6\,ML \ch{WSe2} flakes without encapsulation. All of the measurements were performed at 5\,K. Transfer matrix method simulations for variable distance between the DBR mirrors: (g)~1\,ML, (h)~2\,ML, (i)~3\,ML, (j)~4\,ML, (k)~5\,ML, (l)~6\,ML.}
\label{im:Fig3noBN}
\end{figure*}

\begin{figure*}
\centering
\begin{minipage}[t]{.49\textwidth}
\includegraphics[width=\textwidth]{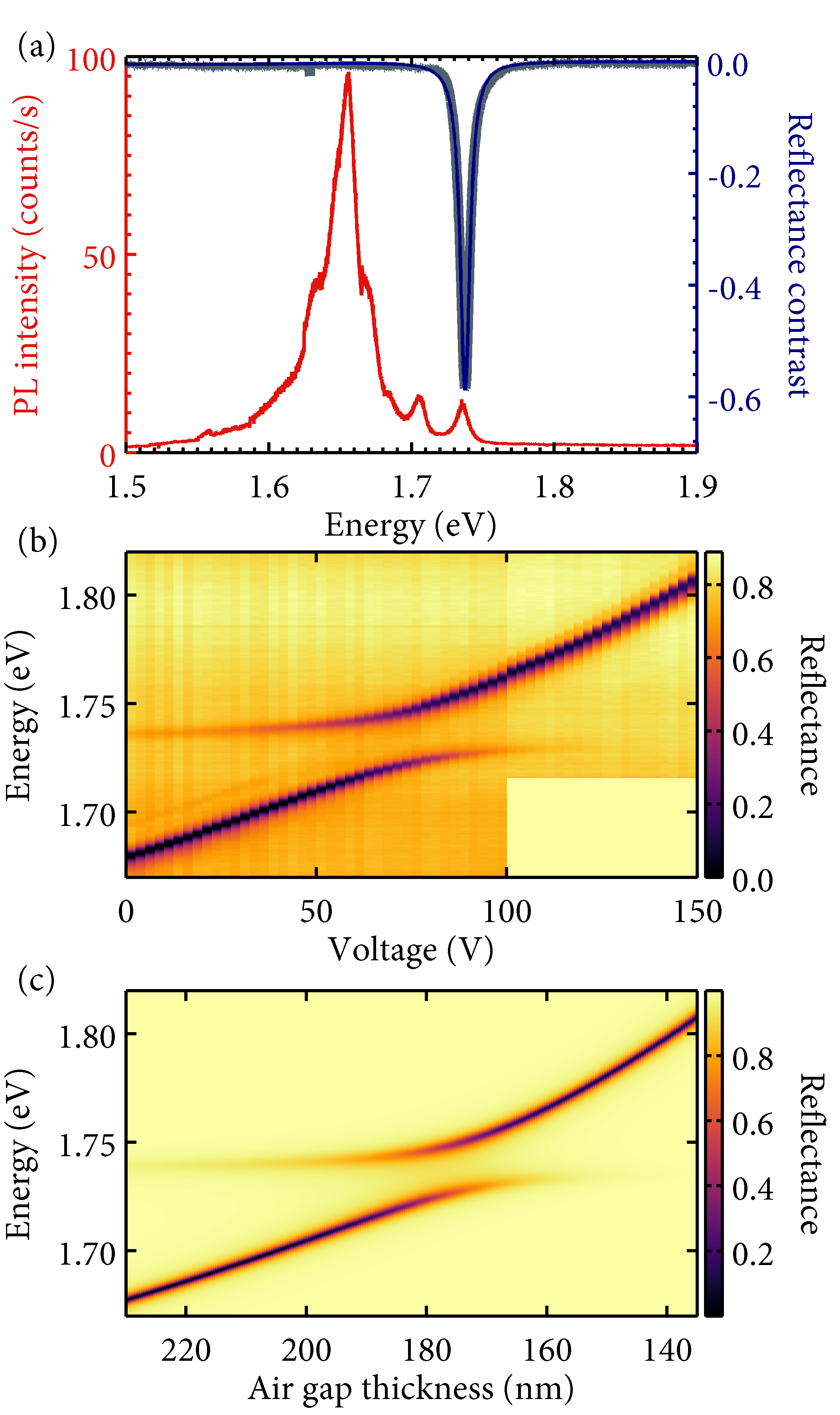}
\caption{\ch{WSe2} monolayer deposited on hBN. (a)~Photoluminescence (red) and reflectance contrast spectra (gray). Blue line presents fitted reflectance spectra. (b) Reflectance at normal incidence for variable cavity length. (c) Modeled reflectance map based on the fitting the blue curve in (a).}\label{im:Fig1BN}
\end{minipage}\quad
\begin{minipage}[t]{.49\textwidth}
\vspace*{-39\baselineskip}
\includegraphics[width=\textwidth]{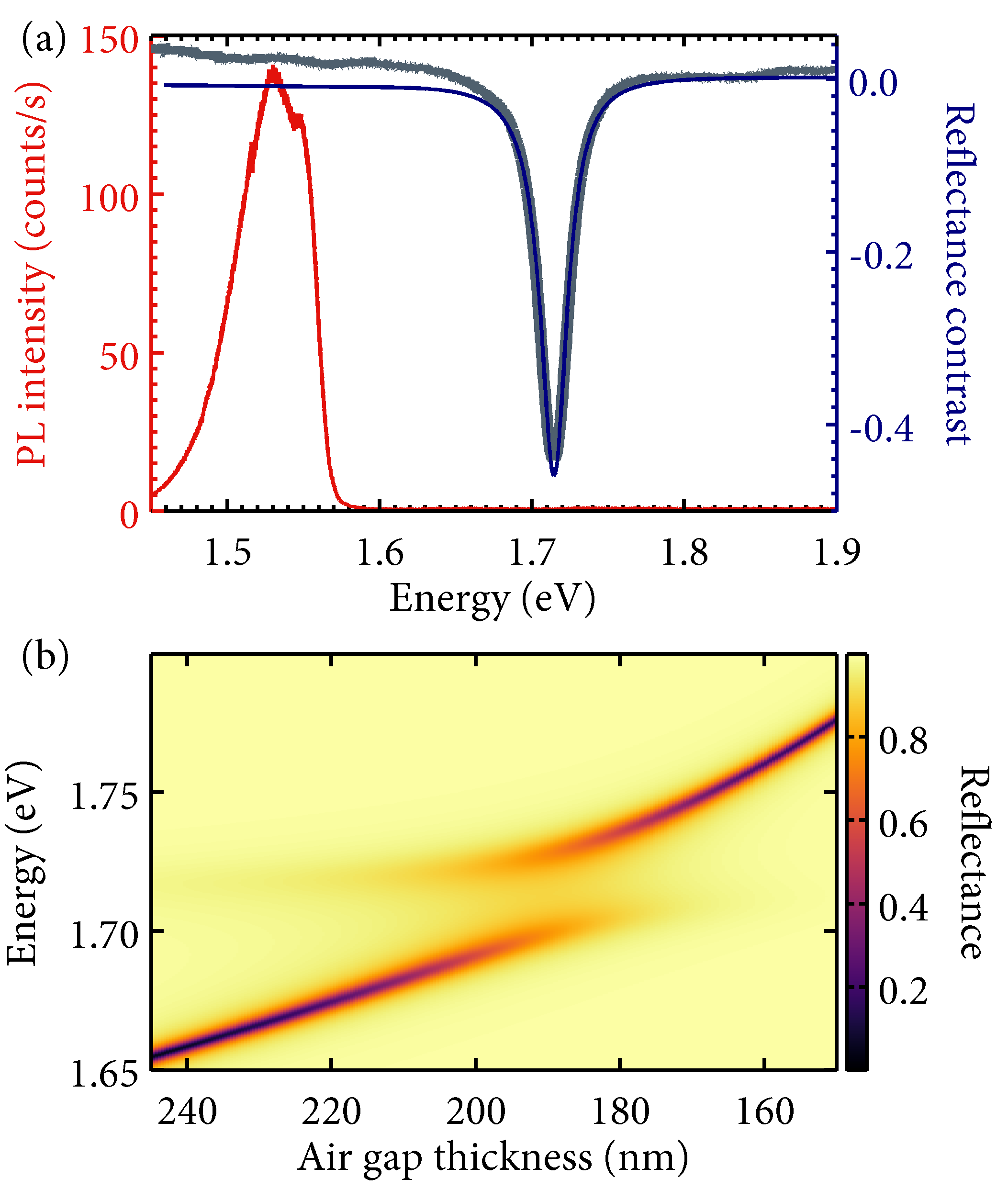}
\caption{\ch{WSe2} bilayer deposited on hBN. (a)~Photoluminescence (red) and reflectance contrast spectra (gray). Blue line presents fitted reflectance spectra.  (b) Modeled reflectance map based on the fitting the blue curve in (a).}\label{im:Fig2BN}

\end{minipage}

\vspace*{5\baselineskip}

\begin{minipage}[t]{\textwidth}
\captionof{table}{Fitting parameters for \ch{WSe2} deposited on hBN with modelled coupling strength in dielectric cavity compared with the values obtained experimentally.}
\label{Table1}
\begin{ruledtabular}
\begin{tabular}{lcccccc}

\textrm{} &	\multicolumn{1}{c}{\textrm{$E_{1s}$ (eV)}} &	\multicolumn{1}{c}{\textrm{$f_{1s}$ (eV\textsuperscript{2})}}	& \multicolumn{1}{c}{\textrm{$\gamma_{1s}$ (meV)}}& \hspace*{0.05cm} &  \multicolumn{1}{c}{\textrm{$\Omega_{1s}^{\rm model}$ (meV)}} & \multicolumn{1}{c}{\textrm{$\Omega_{1s}^{\rm exp}$ (meV)}}  \\
\hline
\textrm{1 ML}   & 1.7366 & 1.10  & ~7.6 & & 21.7 & 24.5 \\
\textrm{2 ML}   & 1.7132 & 1.03  & 20.5 & & 27.6 &      \\
\end{tabular}
\end{ruledtabular}
\end{minipage}
\end{figure*}

\begin{figure*}
\centering
\begin{minipage}[b]{.49\textwidth}
\includegraphics[width=\textwidth]{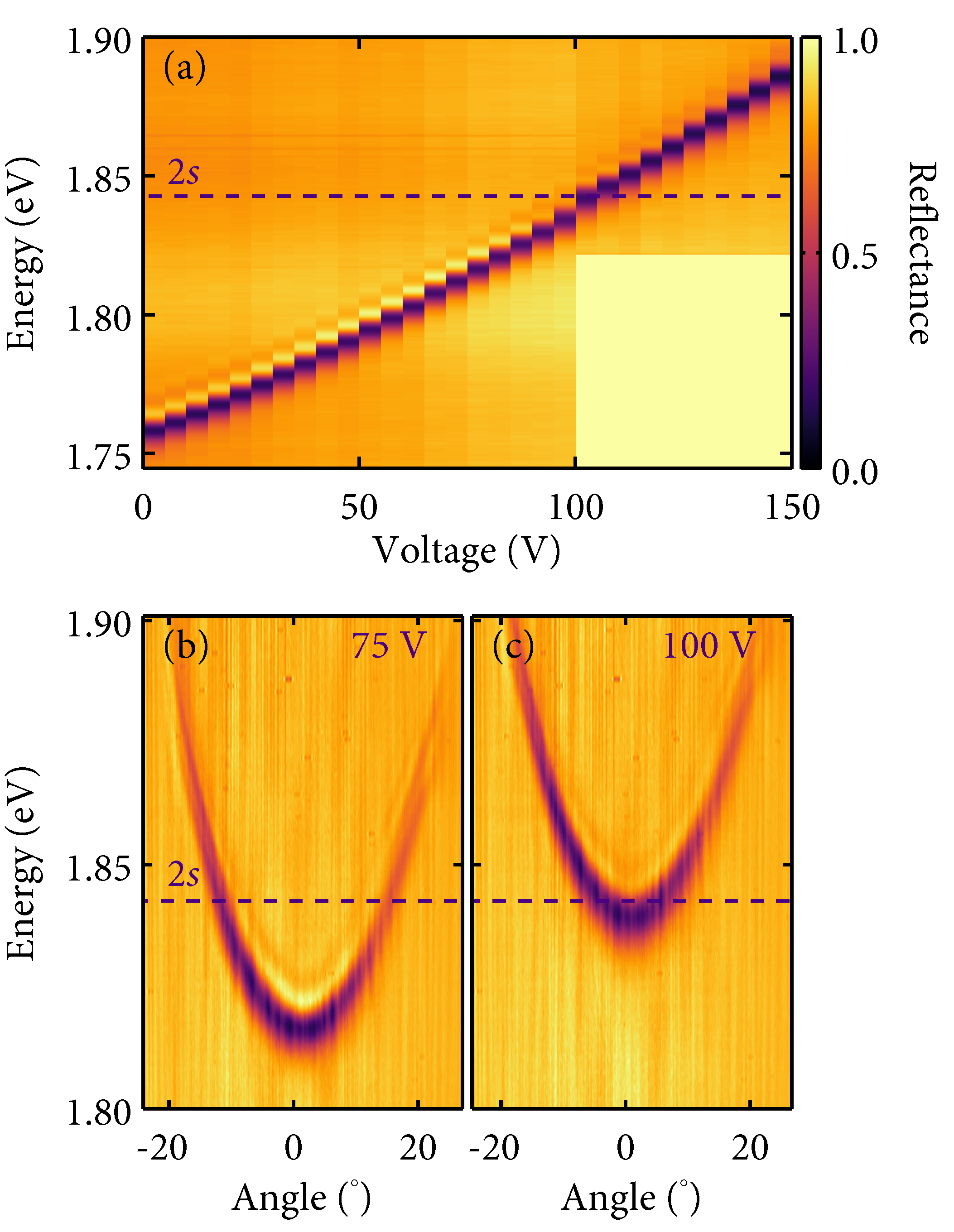}
\caption{Cavity resonant with 2\textit{s} exciton transition in monolayer \ch{WSe2}. (a) Reflectance at normal incidence with cavity mode crossing 2\textit{s} exciton energy. Angle-resolved reflectance spectra at (b)~75\,V and (c) 100\,V applied to the piezoelectric chip. No sign of coupling at 2\textit{s} state energy is visible. }\label{im:Fig2s}
\end{minipage}\quad
\begin{minipage}[b]{.49\textwidth}
\centering
\includegraphics[width=\textwidth]{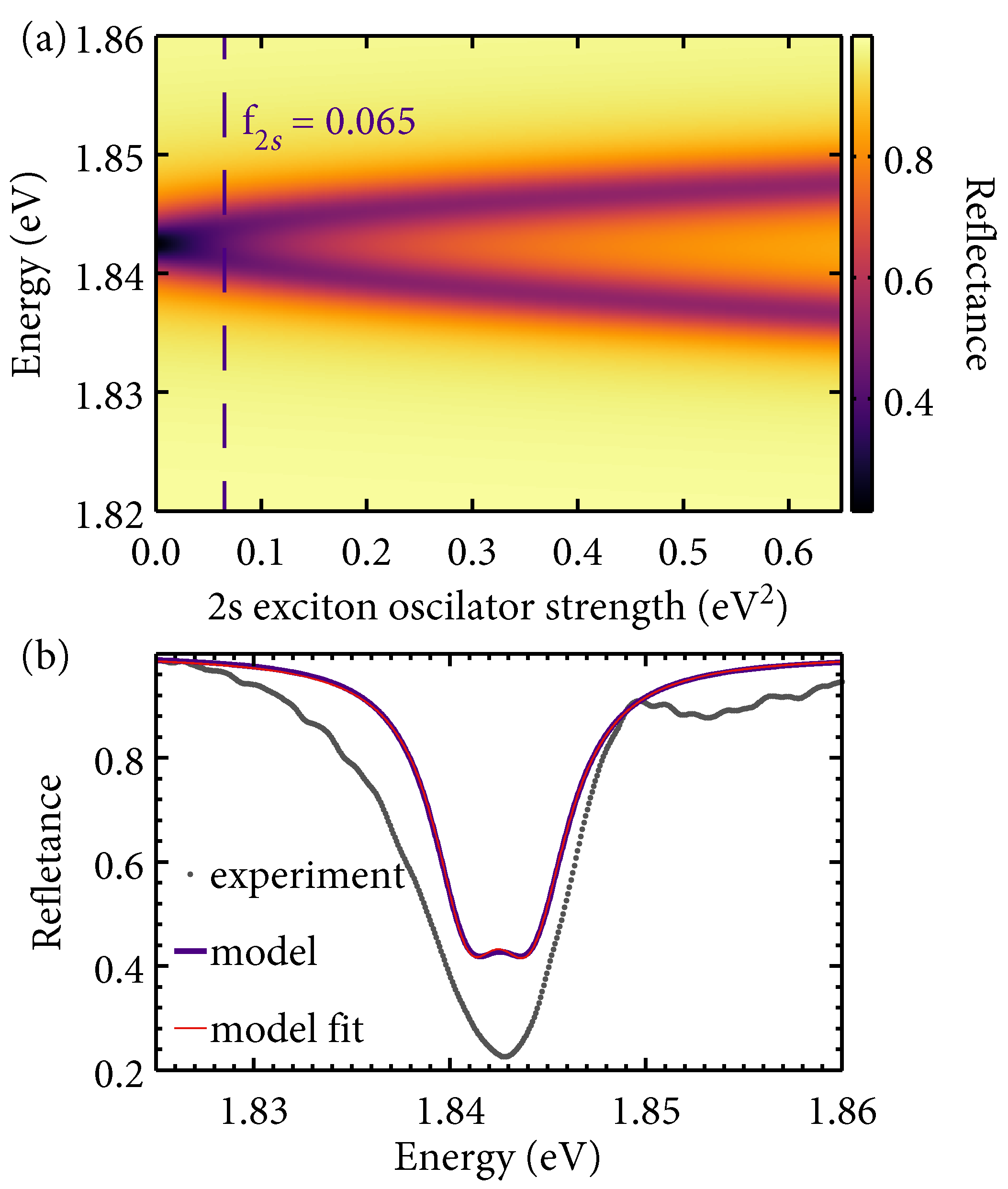}
\caption{Simulations for cavity resonant with 2\textit{s} exciton transition in monolayer \ch{WSe2}. (a) Reflectance at normal incidence calculated for various oscillator strength of 2\textit{s} exciton $f_{2s}$. (b)~Cross section through (a) for experimental value of $f_{2s}$. Sum of two Lorentz functions was fitted to the spectra giving distance between two modes of 3.2\,meV, which is lower than linewidth of 4.5\,meV.  }\label{im:Fig2sModel}
\end{minipage}
\end{figure*}

%
%

%